\UseRawInputEncoding

\documentclass{article}
\usepackage{arxiv}

\usepackage{graphicx,amsfonts,amssymb,amsmath,mathrsfs,hyperref,bm,dsfont}
\usepackage[normalem]{ulem}
\usepackage{amsfonts}  
\usepackage{amsmath}
\usepackage{mathtools}
\usepackage{xcolor}         
\usepackage{array} 
\newcolumntype{H}{>{\iffalse}c<{\fi}@{}}
\usepackage{lipsum}
\usepackage{amssymb}
\usepackage{multicol}
\usepackage{graphicx}
\usepackage{bm}
\usepackage{amsthm}
\usepackage{amsmath}
\usepackage{algorithm}
\usepackage{algpseudocode}
\usepackage{amsthm,amssymb,amsmath,bbm}

\newcommand{\E}{\mathbb{E}}
\newcommand{\RR}{\mathbb{R}}

\newcommand{\F}{\mathcal{F}}

\newcommand{\ud}{\,\mathrm{d}}

\usepackage{authblk}

\usepackage{color}

\newif\ifhyper
\hypertrue
\ifhyper
\hypersetup{
   citecolor = {red},
   colorlinks = {true}, 
   linkcolor = {blue},
   urlcolor = {blue} 
}
\fi

\def\be{\begin{equation}}
\def\ee{\end{equation}}
\def\bea{\begin{eqnarray}}
\def\eea{\end{eqnarray}}

\begin{document}




\title{Quantum-Inspired Tensor Neural Networks \\ for Option Pricing}




\author[1,4]{\thanks{Email: raj.patel@multiversecomputing.com}\hspace{0.5mm} Raj G. Patel}
\author[2]{Chia-Wei Hsing}
\author[2]{Serkan Şahin}

\author[1]{Samuel Palmer}
\author[2,5]{Saeed S. Jahromi}
\author[2]{Shivam Sharma}
\author[1,4]{Tomas Dominguez}
\author[1]{Kris Tziritas}
\author[3]{Christophe Michel}
\author[3]{Vincent Porte}
\author[3]{Mustafa Abid}
\author[3]{St\'ephane Aubert}
\author[3]{Pierre Castellani}
\author[1]{Samuel Mugel}
\author[2,5,6]{Rom\'an Orús}

\affil[1]{%
Multiverse Computing, Centre for Social Innovation, 192 Spadina Ave, Suite 509, Toronto, M5T 2C2, Canada} 
\affil[2]{%
Multiverse Computing, Paseo de Miram\'on 170, 20014 San Sebasti\'an, Spain} 
\affil[3]{%
Cr\'edit Agricole, 12, Place des Etats-Unis - CS 70052 - 92547 Montrouge Cedex, France} 
\affil[4]{%
University of Toronto, Toronto, Ontario M5S 2E4, Canada} 
\affil[5]{%
Donostia International Physics Center, Paseo Manuel de Lardizabal 4, E-20018 San Sebasti\'an, Spain}
\affil[6]{%
Ikerbasque Foundation for Science, Maria Diaz de Haro 3, E-48013 Bilbao, Spain}

\maketitle
\begin{abstract}
    Recent advances in deep learning have enabled us to address the curse of dimensionality (COD) by solving problems in higher dimensions. A subset of such approaches of addressing the COD has led us to solving high-dimensional PDEs. This has resulted in opening doors to solving a variety of real-world problems ranging from mathematical finance to stochastic control for industrial applications. Although feasible, these deep learning methods are still constrained by training time and memory. Tackling these shortcomings, Tensor Neural Networks (TNN) demonstrate that they can provide significant parameter savings while attaining the same accuracy as compared to the classical Dense Neural Network (DNN). In addition, we also show how TNN can be trained faster than DNN for the same accuracy. Besides TNN, we also introduce Tensor Network Initializer (TNN Init), a weight initialization scheme that leads to faster convergence with smaller variance for an equivalent parameter count as compared to a DNN. We benchmark TNN and TNN Init by applying them to solve the parabolic PDE associated with the Heston model, which is widely used in financial pricing theory.
\end{abstract}

\section{Introduction}

Partial Differential Equations (PDEs) are an indispensable tool for modeling a variety of problems in quantitative finance. Typical approaches for solving such PDEs, which are mostly parabolic, rely on classical mesh-based numerical methods or Monte Carlo approaches. However, scaling these approaches to higher dimensions has always been a challenge because of their dependency on spatio-temporal grids as well as on a large number of sample paths. As an alternative, recent advancements in deep learning, which leverages their compositional structure, have enabled us to bypass some of these challenges by approximating the unknown solution using Dense Neural Networks (DNNs) \cite{Raissi, Beck_2019, Han_2018, E_2017}. This has opened up a wide array of possibilities in quantitative finance where we now can consider all participating assets without making any provisional assumptions on their correlations. As a result, we can now consider a whole basket of assets thereby allowing us to transform, extend and solve our problem in higher dimensions.

The basic idea of these approaches is to reformulate the high-dimensional PDEs as forward-backward stochastic differential equations (FBSDE) \cite{Cheridito}. The solution of the corresponding FBSDE can be written as a deterministic function of time and the state process. Under suitable regularity assumptions, the FBSDE solution can represent the solution of the underlying parabolic PDE. Efficient methods for approximating the FBSDE solution with a DNN have been put forward recently in Refs. \cite{Raissi-part1,Raissi-part2}. However, in spite of their apparent success, DNN approaches for solving PDEs are computationally expensive and limited by memory \cite{TNN_NIPS, tn_memory, xue13_interspeech}.

\noindent
In this paper, we show how to overcome this problem by combining Tensor Networks (TN) \cite{RomanTN} with DNNs. TNs were originally proposed in physics to efficiently describe strongly-correlated structures. At the fundamental level though, TNs are nothing but efficient descriptions of high-dimensional vectors and operators. Because of this, TNs have emerged as a promising tool in machine learning (ML) and optimization. In particular, TNs have proven successful in ML tasks such as classification \cite{NIPS2016_6211,Stoudenmire_2018,glasser2018supervised,efthymiou2019tensornetwork,bhatia2019matrix,Liu_2019,9058650}, generative modeling \cite{PhysRevX.8.031012,PhysRevB.99.155131,PhysRevB.101.075135} and sequence modeling \cite{bradley2020modeling}. Following the ideation from Refs. \cite{TNN_NIPS, tnn_osedelets, CACIB_1}, we transform a DNN into what we call a Tensor Neural Network (TNN). In doing so, we observe enhanced training performance and reduced memory consumption. To validate the improvement, we perform an exhaustive search across all the DNN with the same number of parameters as the TNN. However, none of them match the performance of the TNN. Our main test bed for benchmarking is the Heston model, widely used in financial pricing theory. 

\noindent
This paper is organized as follows: in Section \ref{sec:tn}, we briefly review the concept of TNN and show how one can incorporate them in a Neural Network (NN). In Section \ref{TNN-INIT}, we demonstrate how the TNN Init weight initialization scheme works. In Section \ref{Problem}, we then formulate the mathematical problem at hand. In Section \ref{sec:nn-math}, we show how a parabolic PDE mapped into SDEs can be solved using Neural Networks. We then present our results for the Heston model in Section \ref{sec:results} on how TNN and TNN Init outperform DNN. Finally, in Section \ref{sec:conclude} we present our conclusions.

\section{Tensorizing Neural Networks}
\label{sec:tn}

\begin{figure}[!htp]
\centering
\includegraphics[scale=0.6]{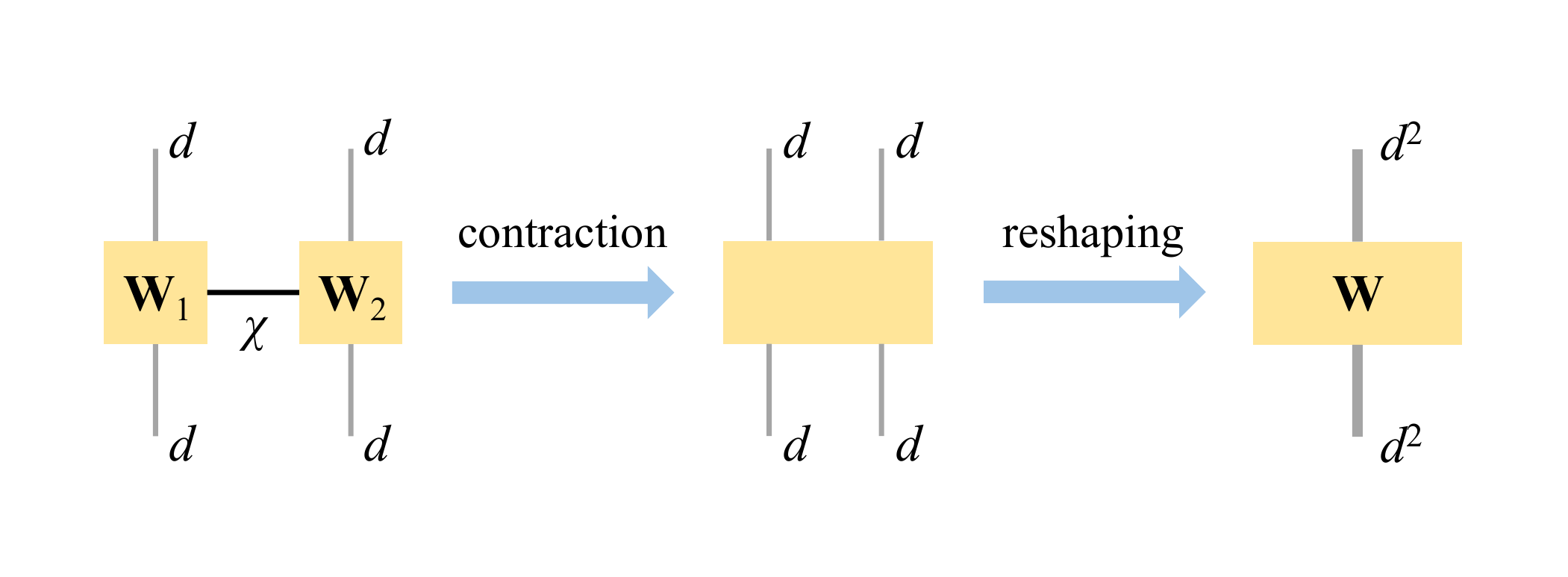}
\caption{The process of contracting a 2-node MPO and reshaping it into the weight matrix $\mathbf{W}$ in each forward pass.}
\label{Fig:Fig3}
\end{figure}

\noindent
A way of tensorizing Neural Networks is to replace the weight matrix of a dense layer by a Tensor Network \cite{RomanTN, CACIB_1}. In particular, we choose an MPO representation \cite{RomanTN} of the weight matrix that is analogous to the Tensor-Train format \cite{TNN_NIPS}, and we call this layer a \emph{TN layer}. This representation, however, is not unique, and is determined by two additional parameters: the MPO bond dimension, and the number of tensors in the MPO. In the simplest case, the MPO may consist of only two tensors, $\mathbf{W_1}$ and $\mathbf{W_2}$, as shown in Fig.~\ref{Fig:Fig3}. The MPO in the figure has bond dimension $\chi$ and physical dimension $d$ as the input and output dimension. The TN layer with such an MPO can be initialized in the same manner as a weight matrix of a dense layer.\newline


\noindent
In the forward pass of the TN layer, we first contract the MPO along the bond index and then reshape the resulting rank-4 tensor into a matrix as shown in Fig.~\ref{Fig:Fig3}. This matrix is the corresponding weight matrix of a TN layer. The weight matrix can then be multiplied with the input vector. We apply an activation function to the resulting output vector, thereby finishing the forward pass. The weight matrix takes the form 
\begin{equation}
    \mathbf{W} = \sum_{\alpha = 1}^{\chi} (\mathbf{A}_\alpha \otimes \mathbf{B}_\alpha), ~~~\mathbf{W} \in \mathbb{R}^{d^2 \times d^2},
\label{sum_of_tensor_product}
\end{equation}

\noindent
where $\mathbf{W}_1 = [\mathbf{A}_1, \mathbf{A}_2, \cdots, \mathbf{A}_{\chi}], \mathbf{A}_\alpha \in \mathbb{R}^{d \times d}$ and $\mathbf{W}_2 = [\mathbf{B}_1, \mathbf{B}_2, \cdots, \mathbf{B}_{\chi}], \mathbf{B}_\alpha \in \mathbb{R}^{d \times d}$ are the two rank-3 weight tensors connected by a virtual bond $\alpha$ of dimension $\chi$. The resulting weight matrix $\mathbf{W}$ is of dimension $d^2 \times d^2$, so it contains $d^4$ elements. Notice that these elements \emph{are not independent} since they come from the TN structure with $2 \chi d^2$ trainable parameters. So, if we initialized the MPO with bond dimension $\chi = d^2/2$, we would have the same number of parameters as a dense layer with $d^2$ neurons. Any choice where $\chi < d^2 / 2$ will result in a weight matrix $\mathbf{W}$ comprising of $d^4 - 2 \chi d^2$ fewer parameters than the weight matrix of a dense layer, thus allowing for potential parameter savings. In principle, when $\chi = d^2$, we have sufficient degrees of freedom to be able to construct an arbitrary $d^2 \times d^2$ matrix. Thus, we expect that by increasing the bond dimension, the TN layer behaves increasingly similar to a dense layer, as shown empirically in \cite{CACIB_1}. 

\noindent
The existence of Kronecker product in Eq.(\ref{sum_of_tensor_product}) implies that there is a correlation between the matrix elements in $\mathbf{W}$, i.e. each element will be a sum of products of elements of the tensors $\mathbf{A}$ and $\mathbf{B}$. The parameters to be trained are not the matrix elements of the weight matrix, but the elements of the individual tensors of the MPO. This can exhibit interesting training behavior and can lead to faster convergence of the loss function as we show in Section \ref{sec:results}.

\noindent
By implementing the TN layer in this way and with a ML library which supports automatic differentiation such as TensorFlow or PyTorch, one can optimize the MPO weights in a similar fashion as those of dense layers in DNN and train the TNN. As an alternative, one could work with an explicit TN layer without contraction of the MPO, including tensor optimizations as in other TN algorithms (e.g., variational), provided one can find a way to decompose the activation function. We observe, however, that for most interesting NN structures, we do not actually need this option. Additionally, the TNN structure is not limited to a single TN layer. It can be extended to any number of TN layers or combinations of dense and TN layers. This provides flexibility when designing TNN architectures which is favorable for the problems of interest.

\section{Tensor Network Initializer}\label{TNN-INIT}
To build upon the structural advantages of TNNs, we propose an alternate framework for initializing the weights. Under this framework, we initialize the weight matrices $\mathbf{W_1}$ and $\mathbf{W_2}$ similarly to how we initialized them while tensorizing neural networks as shown in Fig. \ref{Fig:Fig3}. However, under this scheme, once the MPO is contracted and reshaped into the weight matrix $\mathbf{W}$ during the initialization stage, we only use and update the weights in $\mathbf{W}$ during the learning process. To verify the empirical advantages of this approach, we show preliminary results in Section \ref{results}. 

\section{Problem Formulation}\label{Problem}

\noindent
Although Refs. \cite{Raissi-part1, CACIB_1} have shown results for solving the Black-Scholes-Barenblatt model with a squared payoff, one of the fundamental drawbacks of using the Black-Scholes-Barenblatt model is that it assumes the volatility of asset prices remains constant over time; however, it appears that volatility itself is a stochastic process. To overcome this issue, one of the widely adopted alternatives to the Black-Scholes model is a stochastic volatility model known as the \textit{Heston model}. \newline

Under this model, we aim to solve the following PDE:
\begin{equation}\label{eqn: PDE_CF}
\left\{\begin{aligned}
\frac{\partial u}{\partial t}+\left(r-\frac{1}{2}v\right)\frac{\partial u}{\partial x}+\kappa(\theta-v)\frac{\partial u}{\partial v}+\frac{1}{2}v\frac{\partial^2u}{\partial x^2}+\frac{1}{2}\eta^2v\frac{\partial^2 u}{\partial v^2}+\rho \eta v\frac{\partial^2u}{\partial x\partial v}&=ru,\\
u(T,x,v)&=\phi(x,v).
\end{aligned}\right.
\end{equation}

\noindent
where $\kappa$, $\theta ,\eta>0$ are positive constants, $r\in \RR$ is an interest rate, $v$ is the variance factor of the asset and $u$ is the price process of the contingent claim with a payoff function $\phi$ at time $T>0$. 

Using the Feynman-Kac formalism, this can be re-casted into a system of forward-backward stochastic differential equations. For the forward system, we have two one-dimensional $\mathbb{Q}$-Brownian motions $W^S$ and $W^v$ with correlation $\rho$, and an asset price process $S^c=(S_t^c)_{t\geq 0}$ which satisfies the stochastic differential equation for the Heston model \newline
\begin{equation}\label{eqn: Heston model}
\left\{\begin{aligned}
\ud S_t^c&=rS_t^c\ud t + S_t^c\sqrt{v_t^c}\ud W_t^{S}\\
\ud v_t^c&=\kappa(\theta-v_t^c)\ud t+\eta\sqrt{v_t^c}\ud W_t^{v}.
\end{aligned}\right.
\end{equation}

Here, we add a $c$ as the superscript to denote the continuous nature of the equation which we later drop when we perform discretization. In this model, the volatility of the asset $S$ is given by the stochastic process $(\sqrt{v_t^c})_{t\geq 0}$. Knowing that the underlying is a positive stochastic process, we introduce the semi-martingale $X^c=\log(S^c)$. Through a simple application of Ito's lemma we find that
\begin{equation}\label{eqn: log(S)}
\ud X_t^c=\sqrt{v_t^c}\ud W_t^{S}+\left(r-\frac{1}{2}v_t^c\right)\ud t.
\end{equation}
Comparing Eq. \ref{eqn: log(S)} and Eq. \ref{eqn: Heston model}, we immediately observe the advantage of working with $X$ as opposed to $S$. The absence of the state variable $X$ on the right-hand side of the SDE describing its evolution leads us to fewer numerical errors than directly simulating $S$.

\noindent
Following this, we consider a partition $\{t_n\}_{n=0}^{N\mathrm{d}t}$ of the interval $[0,T]$ with $\Delta t_n = t_{n+1}-t_n$, and we introduce a $\mathbb{Q}$-Brownian motion $Z$ independent of $W^X$ with  
\begin{equation*}
{W_{t}^v\stackrel{d}{\substack{=\\ \mathbb{Q}}} \rho W_{t}^X + \sqrt{1-\rho^2} Z_t}
\end{equation*}
for each $t>0$. If we slightly abuse notation and write $Y_n=Y_{t_n}$ for any continuous stochastic process $Y$ on $[0,T]$, then a simple application of the Euler discretization yields
\begin{equation*}
\left\{\begin{aligned}
X_{n+1}&=X_{{n}}+\sqrt{v_{{n}}}\Delta W_{n}^X + \left(r-\frac{1}{2}v_{{n}}\right) \Delta t_n\\
v_{n+1}&=v_{{n}}+\kappa(\theta-v_{{n}})\Delta t_n+\eta \sqrt{ v_{{n}}}\Delta W_{n}^v,
\end{aligned}\right.
\end{equation*}\\
where $\smash{\Delta W_{n}^X} = W_{n+1}^X - W_{n}^X$,  $\smash{\Delta W_{n}^v} = W_{n+1}^v - W_{n}^v$ and $\smash{\Delta Z_n} = Z_{n+1} - Z_{n}$. Given that we now use discretized versions of Eq. \ref{eqn: Heston model} and Eq. , \ref{eqn: log(S)}, we drop the $c$ indicating the continuous nature of the evolution. To overcome the issue of having an ill-defined square root of $v_{n}$, we truncate $v_{n}$ when it becomes negative leading us to:
\begin{equation}\label{eqn: Euler discretization}
\left\{\begin{aligned}
X_{n+1}&=X_{{n}}+\sqrt{v_{{n}}^+}\Delta W_{n}^X + \left(r-\frac{1}{2}v_{{n}}^+\right) \Delta t_n\\
v_{n+1}&=v_{{n}}+\kappa(\theta-v_{{n}}^+)\Delta t_n+\eta \sqrt{ v_{{n}}^+}\Delta W_{n}^v,
\end{aligned}\right.
\end{equation}\\
where $v_{{n}}^+ := \max\{v_{{n}},0\}$.


\noindent
Once we have the simulations, we use Feynman-Kac's formalism and Ito's Lemma to get the backward SDE of the option which is as follows:

\begin{equation}\label{eqn: BSDE}
\ud u_t^c=r u_t^c \ud t + \begin{bmatrix}
\frac{\partial u_t^c}{\partial X_t^c}\ \frac{\partial u_t^c}{\partial v_t^c}\end{bmatrix}\begin{bmatrix}
\sqrt{v_t^c}&0\\
\eta\rho\sqrt{v_t^c}&\eta\sqrt{1-\rho^2}\sqrt{v_t^c}
\end{bmatrix}\begin{bmatrix}
\ud W_t^{X}\\
\ud Z_t\end{bmatrix}
\end{equation}

subject to the terminal condition $u_T = \phi(X_T,v_T)$, where $c$ refers to the continuous nature of Eq. \ref{eqn: BSDE}. Specifically, to price a European Call option with strike $K$, we would simulate Eq. \ref{eqn: BSDE} with terminal condition $\phi(X_T,v_T) = \max(e^{X_T}-K,0)$. Similarly, to price a European Put option with strike $K$, we would simulate Eq. \ref{eqn: BSDE} with terminal condition $\phi(X_T,v_T) = \max(K-e^{X_T},0)$. To benchmark our approach, we use an analytical solution obtained through Fourier transformation techniques presented in \cite{Duffie}. This Fourier representation depends on the value $u_\omega(t,X_t,v_t)$ of the contingent claim paying $\phi(X_T, v_T) = e^{i\omega T}$ for $\omega \in \mathbb{R}$, which may be computed using approaches presented in \cite{Carr, heston_fourier,mg_stuart_ord_1994} to be 

$$u_\omega(t,X_t,v_t)=e^{A_\omega(t)+B_\omega(t)v_t+C_\omega(t)X_t},$$

\noindent
for

\begin{align*}
A_\omega(t)&=\kappa \theta\left[r_-(T-t)-\frac{2}{\eta^2}\log\left(\frac{1-r_-r_+^{-1}e^{-\gamma(T-t)}}{1-r_-r_+^{-1}}\right)\right]\\
B_\omega(t)&=r_-\frac{1-e^{-\gamma(T-t)}}{1-r_-r_+^{-1}e^{-\gamma(T-t)}},\\
C_\omega(t)&=i\omega,
\end{align*}
where $r_\pm:=\frac{\beta\pm \gamma}{\eta^2}$, $\alpha:=\frac{1}{2}(\omega^2+i\omega)$, $\beta=\kappa-\rho\eta i\omega$ and $\gamma:=\sqrt{\beta^2+2\eta^2\alpha}$. The primary reason for selecting this problem was to benchmark our approach and test its robustness against a model with an analytical solution. However, the utility of our approach would be more evident for models where performing the integrals to obtain $u_\omega$ is not trivial, and the numerical approaches would succumb to a range of difficulties. Under such circumstances, our approach is indeed robust against varying model complexities thereby equipping us with an interesting tool to price more complex and exotic products reliably.

\section{Neural Networks for PDE} \label{NN-Math}
\label{sec:nn-math}

To begin with, we use an Euler discretization with $N$ time steps to simulate $M$ paths, $(X^j, v^j)_{j\leq M}$ of $(X, v)$. Each path is a vector of size $N$,
\begin{equation}
(X^j, v^j) = (X^j_i, v^j_i)_{i\leq N},
\end{equation}
with $(X^j_i, v^j_i)$ corresponding to the value of $(X,v)$ at the $i^{\mathrm{th}}$ point, $t_i$, in the partition of the interval $[0,T]$. The simulation step has resulted in $NM$ vectors of triples $(X^j_i,v^j_i,t_i)$. These will be the inputs of a Neural Network whose task is to learn the value function,

\begin{equation}
u_t=\E\big[e^{-r(T-t)}(X_T-K)_+|\F_t\big],
\end{equation}
at each point $t_i \in [0,T]$, under the base set of parameters as specified in Sec. \ref{sec:results}. The Neural Network outputs $M$ paths,
\begin{equation}
\widehat{u}^j(\theta)=\big(\widehat{u}^j_i(\theta)\big)_{i\leq N},
\end{equation}
with $\widehat{u}^j_i(\theta)$ being an estimator for $u_{i}$. Following this, the output $\widehat{u}^j$ of the Neural Network can be used to estimate the gradient,
\begin{equation}\label{e.cheyette.grad.V}
\nabla_{(X,v)}\widehat{u}^j,
\end{equation}
by automatic differentiation. 

We can now use the estimate \eqref{e.cheyette.grad.V} of the derivative term obtained from the Neural Network to obtain another estimate for the value process $u$. Indeed, an Euler discretization motivates the definition of the estimate $\widetilde{u}^j(\theta)=(\widetilde{u}^j_i(\theta))_{i\leq N}$ with $\widetilde{u}^j_0(\theta)=\widehat{u}^j_0(\theta)$ and
\begin{equation}
\widetilde{u}^j_{i+1}= (1 + r \Delta t) \widehat{u}_i^j + \begin{bmatrix}
\frac{\partial \widehat{u}_i^j}{\partial X_i}\ \frac{\partial \widehat{u}_i^j}{\partial v_i}\end{bmatrix}\begin{bmatrix}
\sqrt{v_i}&0\\
\eta\rho\sqrt{v_i}&\eta\sqrt{1-\rho^2}\sqrt{v_i}
\end{bmatrix}\begin{bmatrix}
\Delta W_i^X\\
\Delta Z_i\end{bmatrix}
\end{equation}
for $1\leq i\leq N$. 

For each batch, we have now defined two estimators for the value process at points in the partition of the interval $[0,T]$,
\begin{equation}
\widehat{u}^j(\theta) \quad \text{and} \quad \widetilde{u}^j(\theta).
\end{equation}
We train the Neural Network by comparing these two estimators to one another, and also by comparing their terminal conditions to the one found using the payoff function $\phi$ and the simulation of $(X, v)$. In other words, we strive to minimize the loss
\begin{equation}
\mathcal{L}(\theta)=\sum_{j=1}^M \sum_{i=1}^N \big(\widehat{u}_i^j(\theta)-\widetilde{u}_i^j(\theta)\big)^2+\sum_{j=1}^M \big(\widehat{u}_N^j(\theta)-\phi(X_N)\big)^2
\end{equation}
Notice that we could have included a term comparing $\widetilde{u}_N^j(\theta)$ and $\phi(X_N)$, but this is not necessary as the triangle inequality implies that this difference is already dominated by the loss function.

\noindent
The goal was to find the solution of a given PDE and we aimed to approximate it using a neural network. In the process of doing so, we first had the PDE alongside an SDE of an underlying state with known initial condition. From this, we derived the SDE of the state variable modeled by the PDE using the Feynman-Kac formalism and a simple application of Ito's Lemma. This SDE has a known terminal condition. Once we have a system of FBSDE, we can use the approach described here to find or approximate the solution of the SDE which, in turn, can be used as an approximation for the solution of the PDE we are interested in. Having seen how learning would take place, we now look into the results about TNN and TNN Init by showcasing their advantages over DNN.

\section{Results: Option Pricing}\label{sec:results}

We can now apply the methodology described in Sec. \ref{NN-Math} to benchmark the performance of DNN. Whereas to benchmark the TNN performance, we apply the methodology described in Sec. \ref{sec:tn} alongside the learning framework described in Sec. \ref{NN-Math}.

\noindent
For the model, unless otherwise specified, we use the base set of parameters: $r=0$, $S_0 = 1$, $\sqrt{v_0} = 20\%$, $\kappa = 3$, $\sqrt{\theta} = 40\%$, $\eta = 1$ and $\rho = -0.5$. Furthermore, $Y_t$ and $Z_t$ represent $u(t,X_t)$ and $Du(t,X_t)$. We partition the time domain $[0, T]$ into $N = 500$ equally spaced intervals. For the loss, instead of using the mean squared error (MSE) which is classically used for regression problems, we use the log-cosh loss which helps in speeding up the convergence as compared to the work in Ref. \cite{Raissi} and which is of the form $\frac{1}{N}\sum_{i=1}^N \ln(\cosh(\hat{y_i} - y_i))$ \cite{CACIB_1}, where $\hat{y_i}$ is the NN estimate and $y_i$ is the target. To optimize the model weights, we use the Adam Optimizer with batch of size 100 and a fixed learning rate of $10^{-3}$. Given the simplicity of the payoff structure, for all our experiments, we use a 2-hidden layer architecture. For simplicity, we only construct TN layers that are symmetric in each input and each output dimension. In practice, we choose the first layer in our NN to be a dense layer with neurons that match the input shape of the second TN layer. That is, a DNN$(x,y)$ corresponds to a two-layer dense network with $x$ neurons in layer 1 and $y$ neurons in layer 2. On the other hand, a TNN$(x)$ corresponds to a two layer TNN architecture with the first being a dense layer with $x$ neurons and the second layer a TN layer with $x$ neurons.

\subsection{Results}\label{results}
\subsubsection{Dense Neural Network vs Tensor Neural Network}

\noindent
We now investigate the behavior of the loss and the option price at $t = 0$ for three different architectures, TNN(16), DNN(4,24) and DNN(16,16). Note that, in comparison with TNN(16), DNN(16,16) has the same number of neurons but more parameters. Whereas, DNN(4,24) has the same number of parameters but a different number of neurons. All three architectures achieve the same accuracy level upon convergence. So, although TNN(16) achieves the same accuracy as DNN(16,16) with fewer parameters, we find DNN(4,24) to be equally good in terms of accuracy and number of parameters. Hence, the number of parameters may not be used as a measure of compression without considering alternative DNN architectures with the same parameter count, which is a major drawback in the experiments performed in Ref. \cite{TNN_NIPS}.
\newline

\begin{figure}[!htp]
    \centering
    \includegraphics[width=8cm, height=5.5cm]{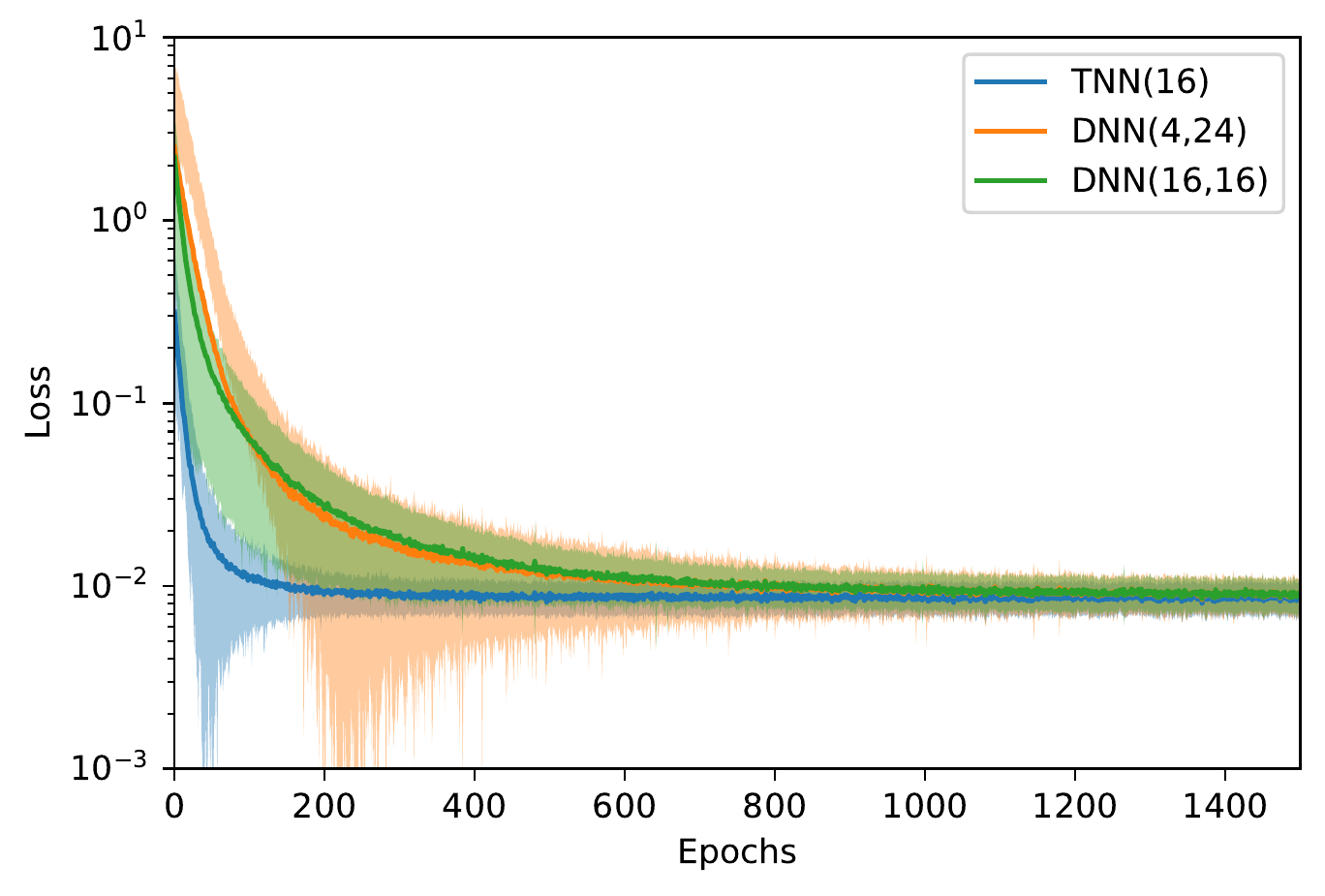}
    \caption{Training loss over epochs for TNN(16) with bond dimension 2 (blue), the corresponding best DNN with equivalent parameters (orange) and the DNN with equivalent neurons (green). The plots illustrates the resulting mean ± standard deviation from 100 runs with different seeds.}
    \label{fig:heston-loss}
\end{figure}

\begin{figure}[!htp]
    \centering
    \includegraphics[width=8cm, height=5.5cm]{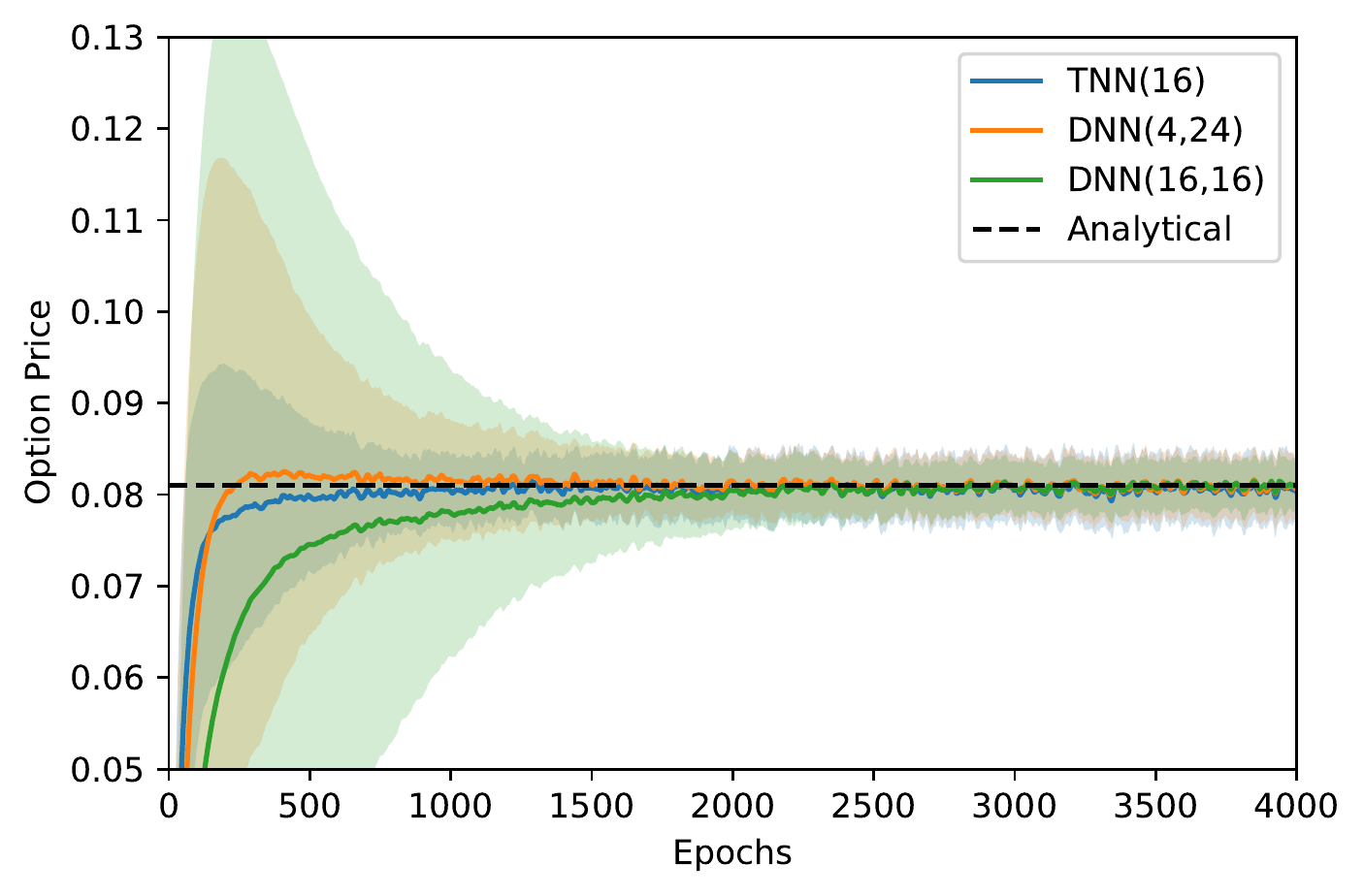}
    \caption{Option Price at $t_0$ over epochs for TNN(16) with bond dimension 2 (blue), the corresponding best DNN with equivalent parameters (orange) and the DNN with equivalent neurons (green). The plots illustrate the resulting mean ± standard deviation from 100 runs. The black dotted line indicates the analytical solution obtained from solving the Heston PDE using Fourier transformation approach discussed at the end of Sec. \ref{Problem}.}
    \label{fig:heston-option-vol}
\end{figure}

\noindent
Moreover, we see in our experiments that the architectures differ in convergence speed. DNN(4,24) converges the fastest among all the DNN architectures with the same parameter count as that of TNN(16). However, we observe that the TNN architecture converges faster than DNN(4,24) as seen in Fig. \ref{fig:heston-loss}. It also converges faster than DNN(16,16). Besides the estimate of the option price as observed in the Fig. \ref{fig:heston-option-vol}, we can also look into the variance of the estimate. As observed in Fig. \ref{fig:heston-option-vol}, we see that TNN not only provides us with better estimates, it also adds stability to our results by reducing the variance as compared to the results of the dense architectures. 

\noindent
In summary, TNN not only allows for memory savings with respect to DNN for the same number of neurons, it also converges faster for the same number of parameters and neurons with a reduction in variance.

\subsubsection{Dense Neural Network vs Tensor Neural Network Initializer}

We now look into the loss behaviour and the option price at $t=0$ for two different architectures, TNN\_INIT(16) and DNN(16,16). Here, TNN\_INIT(16) indicates a two layer dense neural network where the weight matrix is initialized by contracting the MPO and reshaped into the weight matrix $\mathbf{W}$ which is then used during the learning process.

\noindent
In comparison with TNN\_INIT(16), DNN(16,16) has the same number of neurons and parameters which makes it an ideal architecture to compare against for the potential speed-up from using the TN Initializer approach referenced in Sec. \ref{TNN-INIT}. On comparing the two, we note that TNN\_INIT(16) indeed converges significantly faster than DNN(16,16) despite having the same number of parameters and neurons. This also comes with a significant reduction in variance over 100 runs which adds to the increased stability that TNN\_INIT($\cdot$) provides, when compared to a DNN($\cdot$) architecture with an identical number of neurons and parameters and glorot uniform initialization. 

\begin{figure}[htp]
\centering
\includegraphics[scale=0.57]{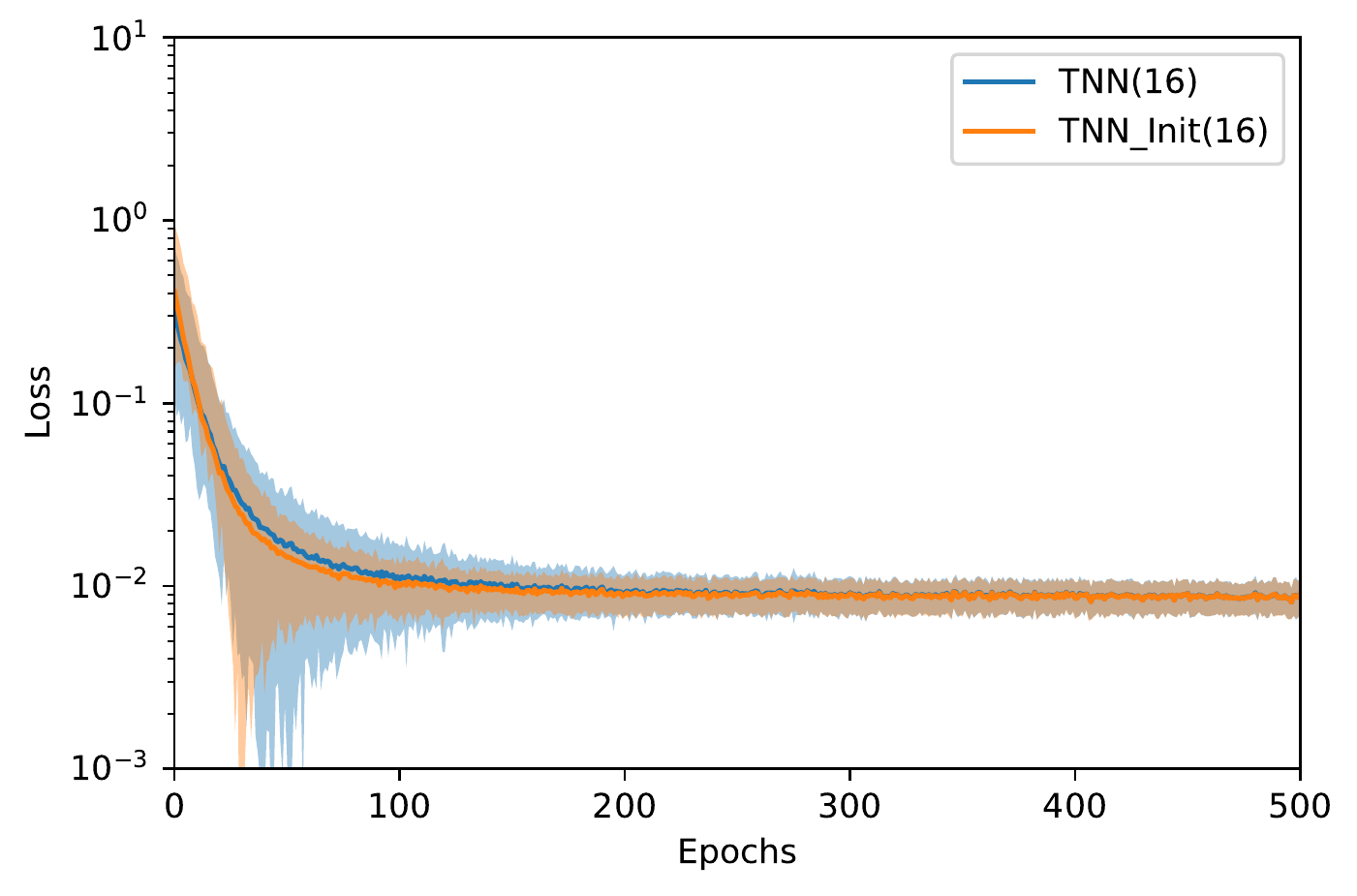} \hfill
\includegraphics[scale=0.57]{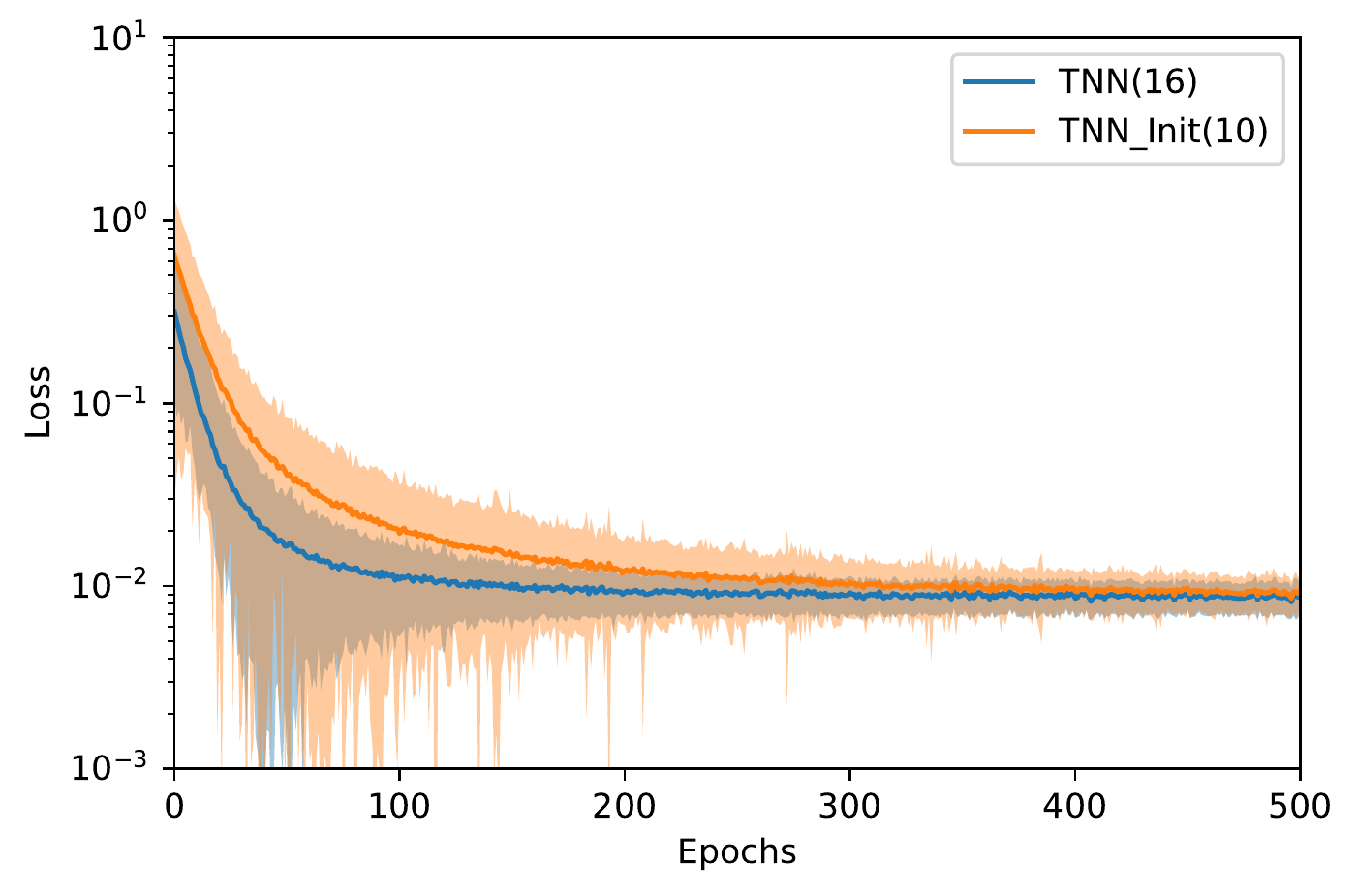}
\caption{(\textbf{left panel}) Training loss over epochs for TNN(16) with a bond dimension 2 and that of a TNN\_Init(16) with a bond dimension 2 (\textbf{right panel}) training loss over epochs for TNN(16) with a bond dimension 2 and that of a TNN\_Init(10) with a bond dimension 2. The plots indicate resulting mean $\pm$ standard deviation from 100 runs.}
\label{Fig:TNN_Init_9_16}
\end{figure}

Given that TNN\_INIT(16) outperforms DNN(16,16), a natural question that arises is how would this fare against the TNN(16) architecture. However, before drawing that comparison, it is essential to know that TNN(16)'s parameter count is 45\% of the parameter count of TNN\_INIT(16). Despite this shrinkage in parameter count, we observe that TNN(16) performs almost identical to TNN\_INIT(16), as shown in the left panel of Fig. \ref{Fig:TNN_Init_9_16}, which adds to the potential of TNNs. To make a fair comparison between the TNN and TNN\_INIT architectures and complete the loop, we compare the performance of TNN(16) and TNN\_INIT(10), which have a similar parameter count. On comparing the two, we indeed see TNN outperforming the  TNN\_INIT architecture, as observed in the right panel of Fig. \ref{Fig:TNN_Init_9_16}, thereby concluding our comparison between the architectures. 

To summarize, we establish that for a similar paramter count, TNN outperforms TNN\_INIT which eventually outperforms the best performing dense architecture with a similar paramter count. Given these advantages, this opens up an interesting avenue for future work within weight initialization for the architectural development of machine learning models.

\begin{figure}[!htp]
    \centering
    \includegraphics[width=8cm, height=5.5cm]{Loss_TNN_Init_Heston_0911.pdf}
    \caption{Training loss over epochs for DNN(16,16) (Orange) and DNN(16, 16) with TN-Initializer (Blue). The plots illustrates the resulting mean ± standard deviation from 100 runs.}
    \label{fig:heston-loss-tnninit}
\end{figure}

\begin{figure}[!htp]
    \centering
    \includegraphics[width=8cm, height=5.5cm]{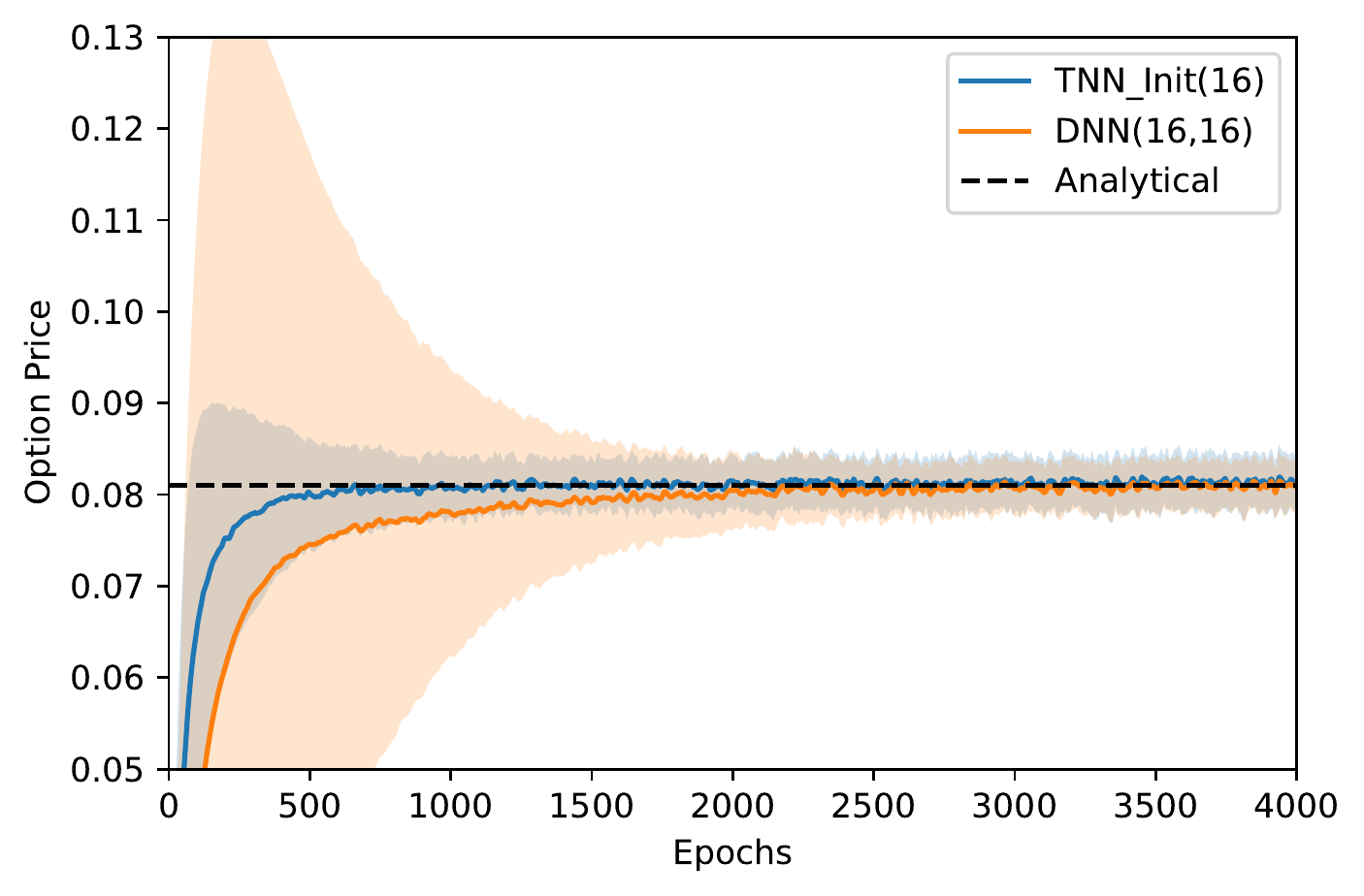}
    \caption{Option Price at $t_0$ over epochs for DNN(16,16) (Orange) and DNN(16, 16) with TN-Initializer (Blue).  The plots illustrates the resulting mean ± standard deviation from 100 runs. The black dotted line indicates the analytical solution obtained from solving the Heston PDE using Fourier transformation approach discussed at the end of Sec. \ref{Problem}.}
    \label{fig:heston-option-vol-tnninit}
\end{figure}

\subsubsection{Runtime Comparison}

\begin{figure}[!htp]
    \centering
    \includegraphics[width=8cm, height=5.5cm]{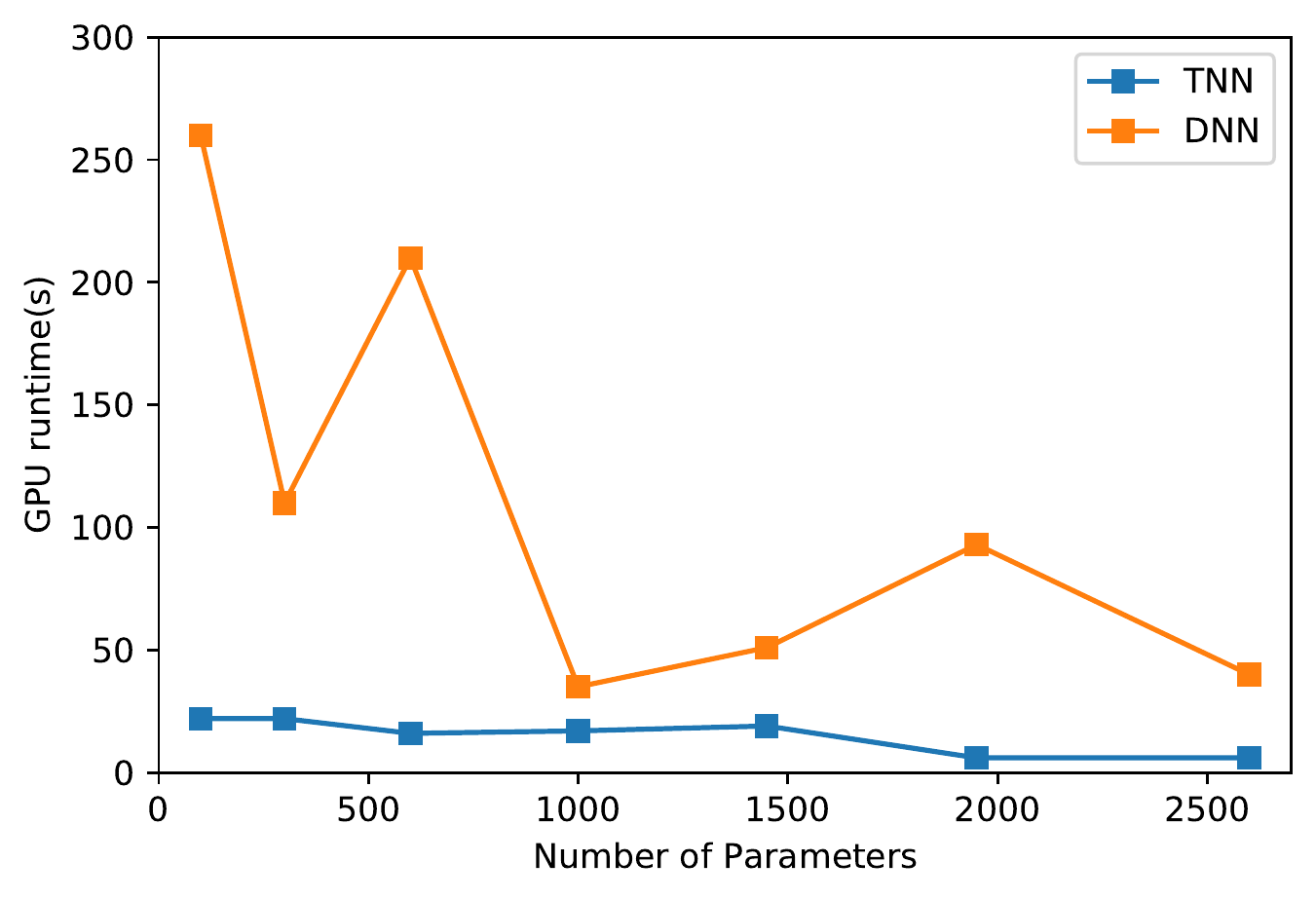}
    \caption{Time taken to converge for a heston model for TNN and DNN}
    \label{fig:heston-runtime}
\end{figure}

\noindent
On comparing runtime for different architectures via AWS EC2 G4dn instances on Nvidia T4 GPU, we observe that all the analysed TNNs converge significantly faster (up to twelve times) than equivalent DNN with same parameter count. Even the smallest TNN architecture ($161$ parameters) outperforms all the analyzed equivalent DNN architectures in terms of convergence speed as shown in Fig. \ref{fig:heston-runtime}. A particular aspect of this result to look into is to understand how this relates to the \emph{compute-efficient training} trade-off. We know that larger models converge to lower errors in fewer gradient updates than smaller models. But furthermore, this increase in convergence speed via fewer gradient updates also outpaces the extra computational costs associated with using larger models \cite{Computeefficiency}, thereby justifying their advantage. This is exactly what we observe here as the runtime to converge goes down as we increase the model size. 

\subsection{Bermudan Options}

We now look into the preliminary results for pricing Bermudan Options using an extension of the classical Longstaff-Schwartz algorithm \cite{l-s-algo}. Under the classical approach, we estimate the conditional expected value using regression. However, the main difficulty there stems from high dimensions when trying to get the conditional expectation. In this section, we demonstrate the results of addressing the issues in the classical model using DNN \cite{lelong} and how we can achieve better results using a TNN. 

\noindent
To understand why neural networks seem to be a good choice over classical regression based approaches, we need to realize that the finite basis functions can not fully represent the conditional payoff whereas we can leverage the approximation power of neural networks. Furthermore, with an increase in dimensionality, we have growing number of basis functions to select from, which eventually results in sub-optimal results, from an accuracy and time perspective, in higher dimensions. Neural networks help here since we are replacing the basis functions and performing gradient descent instead of ordinary least square regression. Instead of choosing the correct basis functions, we learn them, thereby making it easier to scale in higher dimensions. 

Having realized why neural networks are better suited for applications in higher dimensions, we now look at the results for DNN and TNN for pricing Bermudan options in higher dimensions. Here, we price a call option on the maximum of $d$ assets in the Black Scholes model with payoff $(\max_{\substack{i = 1 \dots d}} S^i_T-K)_{+}$ where $d$ is 5 and 10 as investigated in \cite{lelong, RNN}. For parameters, we use $S_0 = \{100, 100, ...\} \in \mathbb{R}^d$, strike ($K$) = 100, dividend = $10\%$, risk-free rate = $5\%$, volatility = $\mathrm{diag}(0.2) \in \mathbb{R}^{d \times d}$, time to maturity = 3 years and $N = 9$ uniform exercise dates. For comparing the performance, we use DNN(16,16) and TNN(16). As noted earlier, in comparison with TNN(16), DNN(16,16) has the same number of
neurons but more parameters. For the loss function, we use the standard mean squared error (MSE). For optimizing the model, we use the Adam Optimizer with a fixed learning rate of $10^{-3}$ with Leaky ReLU as the choice of activation function.

\begin{figure}[htp]
\centering
\includegraphics[scale=0.55]{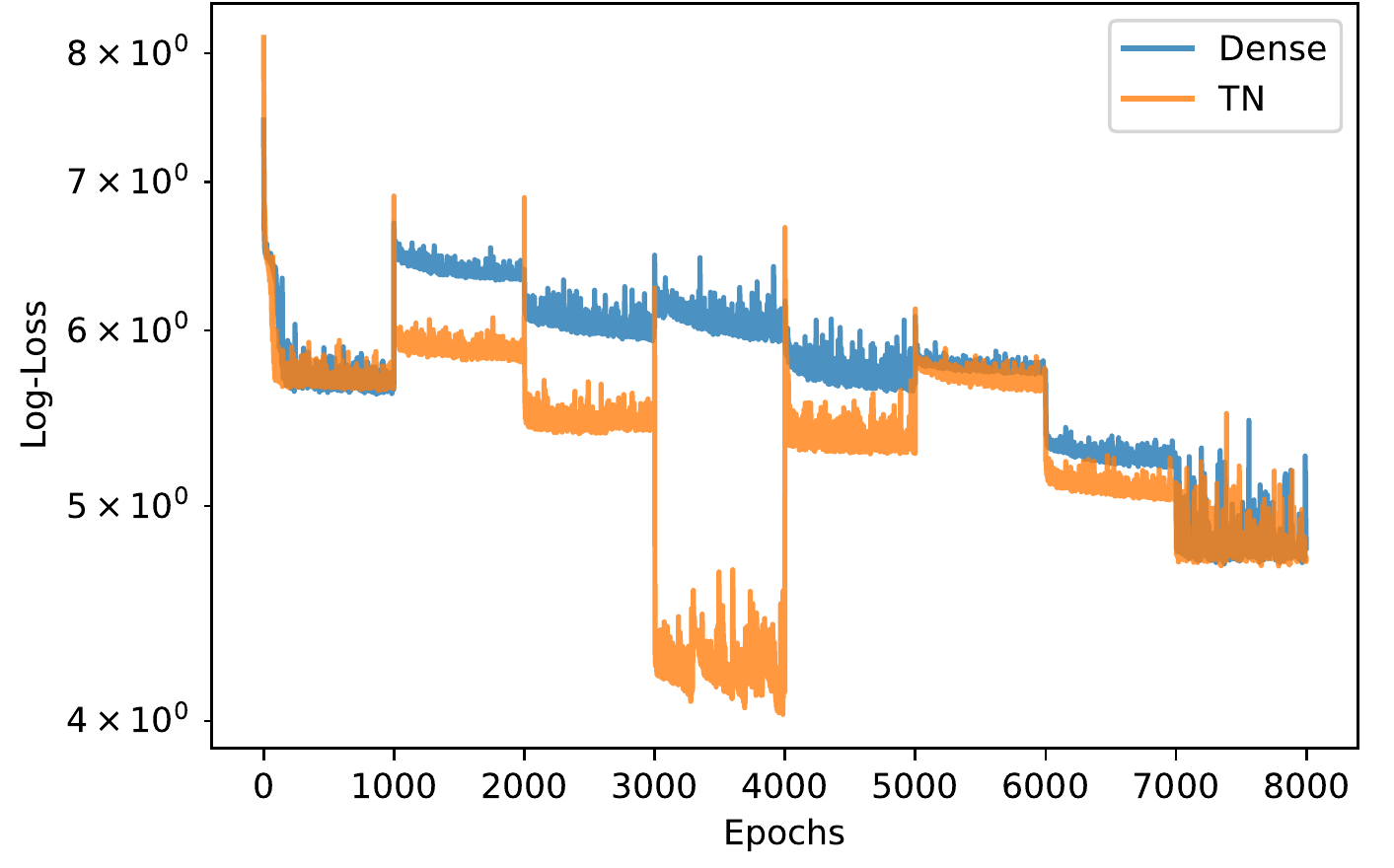} \hfill
\includegraphics[scale=0.55]{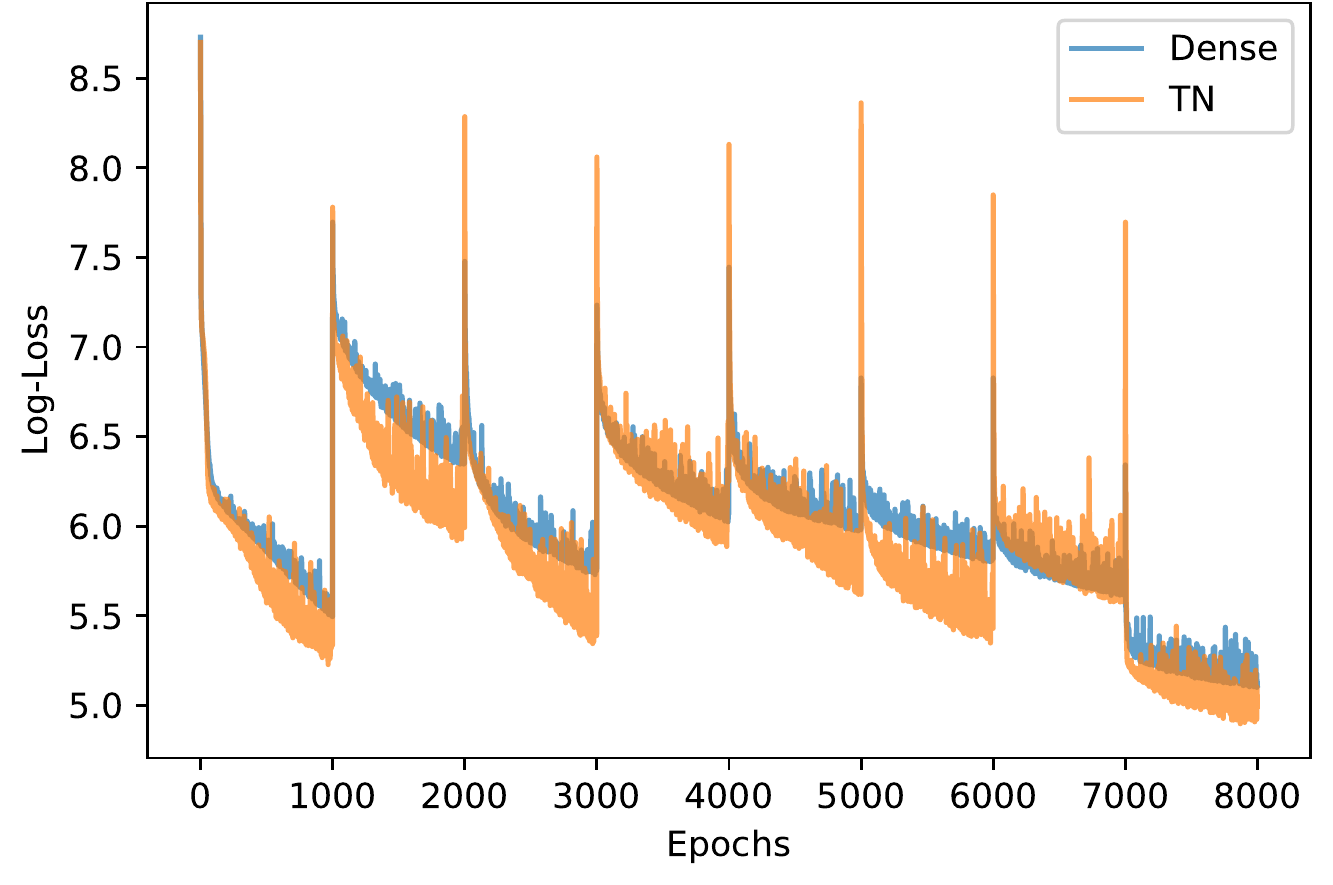}
\caption{Loss evolution for Bermudan Option with (\textbf{left panel}) $d$=5 assets and (\textbf{right panel}) $d$=10 assets}
\label{Fig:Bermudan}
\end{figure}

\noindent
On an elementary level, TNN seems to outperform DNN in terms of minimizing the loss here as observed in Fig. \ref{Fig:Bermudan}. This also corresponds to TNN outperforming DNN in terms of prices (For $d$=5, we get 26.01 $\pm$ 0.18 for DNN and 26.13 $\pm$ 0.18 for TNN. On the other hand, for $d$=10, we get 37.92 $\pm$ 0.27 for DNN and 38.14 $\pm$ 0.25 for TNN). The losses here are based on the performance of TNNs and DNNs at each exercise date. Here, each exercise date is indicated by the discontinuous jumps, i.e., 1000 epochs per exercise date, that we observe in the loss plots. As observed, the loss over epochs for each exercise date is lower for TNNs than DNNs, making them an interesting architecture to investigate. These however are rudimentary results and need more exploration to properly quantify the TNN advantage. 

To tackle this problem rigorously, we can use a sequence of feed-forward neural networks, where each neural network corresponds to learning the continuation value between two exercise dates and the max of continuation value and exercise value as its terminal condition. We can start with terminal time and go back recursively, while learning the continuation value through different neural networks. For each neural network, we can use least squares residual of the associated backward stochastic differential equation as the loss function. An approach like this \cite{Waterloo_American} can facilitate learning the PDE directly rather than using a Monte-Carlo based approach which introduces forward bias. Furthermore, the computational cost of this framework grows quadratically with dimension $d$, in contrast to exponential growth observed in the Longstaff-Schwartz method, which makes it an interesting avenue to look into to compare TNNs and DNNs rigorously.

\section{Conclusions and Outlook}\label{sec:conclude}
We have shown how we can leverage TNN to solve parabolic PDEs for pricing European options. As an extension, we can also tackle PDEs for early exercise options with $N$ discrete time steps by stacking $N$ feedforward neural networks. This can reduce the computational cost from exponential in the Longstaff-Schwartz method to quadratic. However, this is left for future exploration.

Under this regime, we addressed some of the shortcomings in the existing literature when quantifying the advantages of TNN by analyzing parameter savings and convergence speed. Empirically, we demonstrated that TNN provides significant parameter savings as compared to DNN while attaining the same accuracy with a smaller variance. We further illustrated that TNN achieves a speedup in training by comparing TNN against entire families of DNN architectures with similar parameter counts. Besides TNN, we also introduced TNN Initializer, a weight initialization scheme, that empirically outperformed DNN. Despite the absence of theoretical bounds, which can be an area of further analysis, the methodologies described here can be used to improve training from a memory and speed perspective for a wide variety of problems in machine learning. Quantifying the complexity of a problem and adapting the methodology to problems where this approach can provide a significant edge, can be an interesting avenue for future work.

\bigskip 
\noindent
{\bf Acknowledgements -} This work was supported by the European Union’s Horizon Europe research and innovation programme under Grant Agreement No 190145380. We acknowledge the regular fruitful discussions with the technical teams both at Cr\'edit Agricole and Multiverse Computing.

\bibliographystyle{unsrt}
\bibliography{references}

\newpage

\end{document}